# Combinatorial Laser Molecular Beam Epitaxy System Integrated with Specialized Low-temperature Scanning Tunneling Microscopy

Ge He,[1,2] Zhongxu Wei,[1,2] Zhongpei Feng,[1,2] Xiaodong Yu,[1,2] Beiyi Zhu,[1] Li Liu,[1] Kui Jin,[1,2,3,4,*] Jie Yuan,[1,3,4,*] and Qing Huan[1,3,4,5,*]

[1]*Beijing National Laboratory for Condensed Matter Physics, Institute of Physics, Chinese Academy of Sciences, Beijing 100190, China*

[2]*School of Physical Sciences, University of Chinese Academy of Sciences, Beijing 100049, China*

[3]*Songshan Lake Materials Laboratory, Dongguan, Guangdong 523808, China*

[4]*Key Laboratory for Vacuum Physics, University of Chinese Academy of Sciences, Beijing 100190, China*

[5]*CAS Center for Excellence in Topological Quantum Computation, University of Chinese Academy of Sciences, Beijing 100190, China*

(Dated: June 30, 2019)

We present a newly developed facility, comprised of a combinatorial laser molecular beam epitaxy system and an *in-situ* scanning tunneling microscopy (STM). This facility aims at accelerating the materials research in a highly efficient way, by advanced high-throughput film synthesis techniques and subsequent fast characterization of surface morphology and electronic states. Compared with uniform films deposited by conventional methods, the so-called combinatorial thin films will be beneficial to determining the accurate phase diagrams of different materials due to the improved control of parameters such as chemical substitution and sample thickness resulting from a rotary-mask method. A specially designed STM working under low-temperature and ultra-high vacuum conditions is optimized for the characterization of combinatorial thin films, in an XY coarse motion range of 15 mm × 15 mm and with sub-micrometer location precision. The overall configuration as well as some key aspects like sample holder design, scanner head, and sample/tip/target transfer mechanism are described in detail. The performance of the device is demonstrated by synthesizing high-quality superconducting FeSe thin films with gradient thickness, imaging surfaces of highly oriented pyrolytic graphite, Au (111), $Bi_2Sr_2CaCu_2O_{8+\delta}$ (BSCCO) and FeSe. In addition, we have also obtained clean noise spectra of tunneling junctions and the superconducting energy gap of BSCCO. The successful manufacturing of such a facility opens a new window for the next generation of equipment designed for experimental materials research.

* Authors to whom correspondence should be addressed: huanq@iphy.ac.cn, yuanjie@iphy.ac.cn and kuijin@iphy.ac.cn

# I. INTRODUCTION

The emergence of exotic electronic states relies on an exact control of physical parameters, both in synthesis and manipulation[1-4]. For instance, giant magnetoresistance emerges in multilayers composed of alternating ferromagnetic and non-magnetic conductive layers[1]. By doping a Mott insulator, the electronic state evolves from insulating state to high-$T_c$ superconducting state, then to Fermi liquid regime, along with various ordered states or fluctuations[2-4]. Accurate and unified phase diagrams of the materials will be beneficial both, for the comprehension of basic physical concepts, as well as applications. However, traditional methods for exploring new materials are becoming inefficient in view of the rapidly growing demands from various fields. To obtain a complete doping-dependent phase diagram by the conventional one-at-a-time strategy, one must synthesize and characterize samples one doping level after another[3, 5, 6]. As the number of elements in the compounds increases, the workload increases quadratically. As a result, it is nearly impossible to establish a precise phase diagram for compounds comprised of multiple elements. Moreover , it is hard to pin down critical chemical compositions like quantum critical points (QCPs), which are important for the understanding of competing orders in condensed matters[7-9]. Therefore, developing new technologies with both, high-efficient synthesis and characterization becomes vital for materials science.

The combinatorial method first introduced by Hanak in 1960s supplies a high-throughput approach to synthesize samples efficiently[10]. This technique has shown significant advantages for exploring new materials[11] and investigating their physical properties[12]. Up to now, the combinatorial film (combi-film) deposition technique has been through three generations[13]: co-sputtering or co-evaporation[14], array mask technique[15, 16] and combinatorial laser molecular beam epitaxy (Combi-LMBE)[17]. Overcoming the drawbacks of nonlinear and uncontrollable composition gradients in the first- and discrete stoichiometric ratio in second-generation techniques, Combi-LMBE technique could provide one or two-dimensional composition-spread films by a horizontal mobile mask system. Most established Combi-LMBE systems use reciprocating motion of the mask by a precisely controlled motor[17, 18]. In this case, the method induces an accumulative error that eventually has an impact on the accuracy of the composition. In addition, repeated forward and backward operations of the mask will result in the deformation of mechanical components. Clearly, alternative methods realize the composition gradient are needed to overcome the drawbacks mentioned above.

Besides the challenges in synthesis, the lack of proper characterization tools for the combi-film severely limits the wide applications of the high-throughput methods in materials research, where a probe with high spatial resolution is necessary. There are some commercial instruments available, e.g. wavelength dispersive spectroscopy (WDS)[19], x-ray fluorescence spectroscopy (XFS)[20] and x-ray diffractometer[21, 22] for the characterizations of structure and component, and fluorescence spectroscopy[23] and transmission spectroscopy[24] for inspecting optical properties. In addition, some laboratory techniques have been developed suitable for the combi-films, such as scanning superconducting quantum interference device (SSQUID)[25] and scanning Hall microscopy[26, 27] to detect weak magnetic signal, scanning near-field microwave microscopy[28-30] to characterize the dielectric properties, as well as spring probe array[11, 31] to measure the electrical transport properties. To our knowledge, there is still no report on the implementation of local probing of electronic states of combinatorial samples. Scanning tunneling microscopy/spectroscopy (STM/STS) has been widely used in the study of molecular vibration modes[32], spin flip[33], Kondo effect[34], and superconductivity[35], etc. STM probing requires very clean surfaces, which can only be achieved by *in-situ* ion bombing, annealing or cleaving, but most of combinatorial samples cannot satisfy such requirements. Besides, a large range horizontal positioning up to 10 mm × 10 mm in a high precision spatial resolution is required to match the merits of common combi-films. These specs are out of range of most commercial and home-made STM systems with limited XY coarse motion (less than 5 mm) and not suitable for composition spreads[36-39].

There are quite a few of MBE-STM systems that have been implemented previously[40-42]. However, MBE is often used in uniform film deposition in these systems. One can obtain only one doping at one deposition. By contrast, Combi-LMBE is designed to deposit phase spread films in one or two dimensions. In addition, compared with conventional MBE systems, LMBE systems are good at growing oxide samples. Besides, the XY coarse motion of the sample stage in STM scanner heads of these systems is less than 5 mm and without precise positioning, which makes it unsuitable for combi-films characterization.

In this paper, we describe the design, the build, and the performance testing of a newly developed combined system of Combi-LMBE and low-temperature ultra-high vacuum scanning tunneling microscopy (LT-UHV-STM). The system aims at synthesizing combi-films and characterizing their surface morphology and electronic states *in-situ* with high efficiency and precision. Compared with the commercial Combi-LMBE system, we used a rotatable circular mask to avoid the accumulative error from reciprocating motion of the flat mask. Within the STM unit, we designed a scanner head with large XY coarse motion range (15 mm × 15 mm) and a positioning resolution which is better than 1 $\mu$m at liquid helium temperature. In order to test the performance of the system, we deposited FeSe film with gradient thickness by utilizing the rotation mask. The thickness-gradient samples show high quality and nearly linear-in-position thickness ranging from 28 nm to 280 nm, with the superconducting transition temperature between 8 and 10 K. Besides, we obtained high-quality images on highly oriented pyrolytic graphite (HOPG), Au (111), $Bi_2Sr_2CaCu_2O_{8+\delta}$ (BSCCO), and FeSe surfaces, as well as $dI/dV$ spectra of BSCCO. The system exhibits good stability both, with respect to combi-film deposition and STM measurements. Its features should prove especially advantageous in superconductivity research.

## II.    SYSTEM DESIGN

### A.    Ultra-High Vacuum Chamber

Figure 1 shows a 3D model and a photograph of the system. The whole system with a weight of 1.6 ton is supported by a T-shape aluminum alloy frame with a cross section of 80 mm × 80 mm. Three pneumatic vibration isolators are located at the three terminal vertexes of the frame, respectively (see the black cylinders in Fig. 1 (a)). The estimated mass center of the system is shown by a red dot in Fig. 1 (a). The system consists of six chambers, that is, Combi-LMBE chamber, radial telescopic transfer arm (RTTA) chamber (from Kurt J. Lesker Company), STM chamber, preparation chamber, load-lock chamber and buffer chamber (see Fig.1 (a) and (b)). The load-lock chamber is for loading samples, targets, and STM tips. The preparation chamber is installed with a customized manipulator (from UHV Design Ltd) and an ion gun. Cycles of ion bombardment and annealing (up to 1200 ℃) of the sample can be performed in this chamber. The details of the buffer chamber, the Combi-LMBE chamber, RTTA chamber and the STM chamber will be discussed below.

To obtain ultra-high vacuum, a multi-level pump system is used including mechanical pumps, turbo-molecular pumps, sputter-ion pumps, titanium-sublimation pumps (TSP), and non-evaporable getter (NEG) pumps. A 1600 L/s turbo-molecular pump is installed between the RTTA and the Combi-LMBE chambers, which pumps these two chambers as well as the STM chamber. Another 300 L/s turbo-molecular pump, responsible for pumping the preparation and the load-lock chambers, is mounted under the preparation chamber. In order to reach a higher vacuum level, the outlet of these two turbo-molecular pumps is connected to the buffer chamber, with an 80 L/s turbo-molecular pump and a 40 $m^3$/h mechanical pump installed to achieve a vacuum better than $10^{-5}$ Torr. When carrying out the STM characterization, all the turbo-molecular pumps and mechanical pumps are turned off to minimize vibrations. Instead, the UHV environment is maintained by three ion pumps of 400 L/s (integrated with TSP and Cryoshroud), 300 L/s (integrated with NEG) and 300 L/s (integrated with NEG), which are mounted under the Combi-LMBE chamber, the RTTA chamber and the STM chamber, respectively. Three 5 kW heaters are used to bake the system (see the rounded light brown colored parts in Fig. 1 (a)). The whole system is covered by a home-designed heat-insulation tent during the baking. The operating background pressure of the Combi-LMBE chamber, the RTTA chamber and the STM chamber can easily be as low as ~ $10^{-10}$ Torr after one week's baking at 130 ℃.

### B.    Transfer system

As described in the previous section, this is a complex system which has six chambers with multiple functions. Thus, it is crucial to carry out a reliable protocol for a smooth transfer of sample, substrate, tip, and target among different chambers. Figure 2 shows the transfer mechanism of the system. It is composed of one transfer rod, three wobble sticks, three sample manipulators and the RTTA. In this system the RTTA, which has a scalable and rotatable arm to move vertically, is the key element which connects the three main functional chambers, i.e., the preparation chamber, the Combi-LMBE chamber and the STM chamber. Two storage carousels are attached to the end of the transfer rod and the RTTA arm as shown in the inset of Fig. 2. Each storage carousel can accommodate up to eight sample holders and one target.

All the samples/substrates/tips mounted on the holder can be loaded on the storage carousel from the load-lock chamber, and then be further delivered into the preparation chamber by pushing the transfer rod. A wobble stick installed on the preparation chamber is used to move the samples/substrates/tips to the manipulator or the other storage carousel on the RTTA arm. The other two main functional chambers connected to the RTTA chamber have wobble sticks as well. Those wobble sticks fulfil samples/substrates/tips transfer between the RTTA storage carousel and the manipulator or the STM scanner head. The transfer of target is straightforward. As shown in the insert in Figure 2, a fork attached to the front of the storage carousel can carry the target and directly hand it to another fork or the target holder docks on the target stage.

The circumferential arrangement of the three main function chambers around the RTTA chamber presents three advantages: 1) Higher space efficiency, making the whole system more compact; 2) Higher transfer efficiency, so samples/substrates/tips transfer can be realized between any two chambers; 3) Extendibility, permitting new functional chambers to be easily added to the spare ports of the RTTA chamber.

### C. Sample holder and sample holder dock

Both the sample holder and the sample holder dock designs are critical for this system. Special requirements have been considered for extended function. First of all, the sample holder should be compatible with other systems, so that other in-situ characterizations can be done as well. Secondly, the sample holder should be able to bear a wide temperature range (from liquid helium temperature up to 1000 ˚C). Thirdly, physical and chemical properties, including thermal and electrical conductivities, chemical stability in high oxygen environment at high temperature, etc. should be taken into account. Finally, we must position the sample holder precisely in different chambers, so that reliable and reproducible data can be obtained.

We have chosen the popular flag-style shape for the sample holder as can be seen in Fig. 3(a). The sample holder is made up of Inconel 718 which possesses outstanding physical and chemical properties. Samples or substrates are fixed mechanically by a spring plate in the central square area (10 mm × 10 mm). Three spherical grooves on the back of the sample holder are used to locate the position. The sample holder is inserted into a sample holder dock as seen in Fig. 3(b). The dock consists of top and bottom plates, and a BeCu spring. Three position balls are pressed into the corresponding holes of the top plate by the BeCu spring. The exposed parts of the balls on the other side of the top plate can be fitted with the spherical grooves on the back of sample holder for precise positioning.

Considering the functional difference of the sample holder docks in the Combi-LMBE chamber and the STM chamber, we have introduced corresponding alterations as shown in Fig. 3(c) and (d). When we deposit the combi-film, the sample holder needs to be heated up to a high temperature (typically 300-800 ℃). Laser as a stable and efficient heater is widely used in deposition systems[43, 44]. In order to heat the sample holder and measure the temperature, a 10 mm clearance hole has been made to expose the back surface of the sample holder, an 808 nm wavelength laser is mounted on the top of Combi-LMBE chamber and focused on the back of the sample holder with a spot size of 8 mm as shown in Fig. 3(c). The temperature of the sample holder is measured by infrared thermometer, mounted on the Combi-LMBE chamber focusing at an angle of 45 degrees. For the STM sample holder dock (Figure 3(d)), we embed a diode sensor and a 50Ω/1W chip resistor to measure and adjust the sample temperature.

## D. Combi-LMBE

The Combi-LMBE chamber is equipped with three customized UHV stages (from UHV design Ltd.), i.e. sample stage, mask stage and target stage as shown in Fig. 4(a). The sample stage includes four step-motors for X/Y/Z motions and rotation. The mask stage includes two step-motors for linear motion and mask rotation. A home-made big bowl-shape mask is fixed on the mask stage spindle. The rotary motor possesses a resolution of 0.02 degrees, corresponding to 40 $\mu$m per step, which meets the requirement on combi-film deposition[45]. Above the mask stage, the e-gun of reflection high-energy electron diffraction (RHEED) is mounted aiming at the sample center at an angle of 2 degrees deviation from the sample surface to monitor the film growth *in-situ*. The RHEED screen faces the RHEED e-gun to collect the reflected electrons. The target stage includes three step-motors for Z displacement, revolution and rotation. A 6-seat target holder dock is mounted on the revolution plate. A camera is mounted on the bottom center of the Combi-LMBE chamber for assisting in the alignment of the mask and the sample holder. A 248 nm excimer laser (from Coherent GmbH) at an angle of 45 degrees to the target surface is used to ablate the target material during the deposition.

The typical deposition procedure is illustrated in Fig. 4(b): (i) The mask is rotating at constant speed, and the laser starts to shoot and ablate the target A once the right edge of the window in the mask is aligned with right edge of the substrate as shown in ①. (ii) The laser stops shooting until the whole substrate is sheltered by the mask and the first-half period is finished (see ② and ③). (iii) Target B is automatically rotated to the position of target A, and the second-half period starts once the left edge of the window in the mask is aligned with the right edge of the substrate as shown in ④. (iv) The second-half period is continued until the left edge of the substrate and the left edge of the mask overlap (see ⑤ and ⑥). From steps ① to ③, the A component is deposited with roughly linear distribution, from one unit-cell at one end of the substrate to zero coverage at the other end (sketched by blue triangle in Fig. 4(b)). After finishing steps ④ to ⑥, the B component is obtained with a reversed thickness distribution compared to the A component at the same deposition rate and temperature (sketched by red triangle in Fig. 4(b)). Thus, we can obtain a unit-cell thin film with composition continuously varying from A to B. Repeating the above procedure, one can get combi-films of desired thickness. For example, a 100 nm thick $(00l)$-oriented $(La,Ce)CuO_4$ film (the $c$-axis lattice constant ~ 1.2 nm) needs more than 80 periods. Following a similar procedure, one can also get thickness-gradient samples or superlattices by controlling the motion sequence of masks and targets using a home-made LabVIEW program.

## E. STM

The STM chamber is equipped with a commercial bath-type cryostat, a double layer cold room and a home-made scanner. The cryostat consists of two vessels, i.e. a 17 L liquid nitrogen vessel and a 4 L liquid helium vessel. The cold room is mounted at the bottom of the cryostat to isolate from ambient thermal radiation. The STM scanner head is hung inside the 4 K shield by three Inconel springs which provide vibrational isolation from the cryostat. The total weight of the scanning module is 760 g. The resonant frequency of the suspended scanning module is approximately 2.7 Hz. Eddy current damping is provided by 8 SmCo-magnets mounted on the bottom of the scanner head as shown in Fig. 5(a) and (b). The scanner head is thermal−anchored on a gold-plated oxygen-free high-conductivity copper piece at the bottom of the liquid helium vessel by several copper braids as shown in Fig. 5(c). The scanner head can be clamped, thereby providing a good stability for transferring tips and samples and a good thermal contact. The details about cold room, eddy current damping and scanner clamping mechanism can be found in Ref. 42.

Distinguishing the system from traditional STM systems[36-39], the closed-loop XY coarse motion of the sample has a large range. Therefore, the tip can be precisely put on the sample surface for determining its physical and chemical properties like doping level, film thickness, etc. For this, we fixed the STM sample holder dock onto two vertically stacked commercial piezo modules as seen in Fig. 5(a) (ANPx321- closed loop, from Attocube system Inc.). The range of each motor is 15 mm. The fine linear positioning ranges are 0.8 $\mu$m at 4 K and 5 $\mu$m at 300 K respectively. The lower part of the scanner head is a typical Pan-type Z coarse motion with 6 piezo stacks as seen in Fig. 5(b). A simplified inertial tip-approach method is used in this scanner head. More details can be found in Ref. 42.

## III.    PERFORMANCE

### A.    Film deposition and characterization

To test the performance of the Combi-LMBE unit, we grew a FeSe thin film with gradient thickness on a LiF substrate. The background vacuum of the chamber was better than $7.0 \times 10^{-9}$ Torr. The FeSe films were grown with the target-substrate distance of $\sim 50$ mm, laser pulse energy of 350 mJ and repetition rate of 4 Hz, and the substrate temperature of $\sim 350$ ˚C. Before the deposition, the substrate was aligned with one edge of the window in the mask, as explained in section II.D. During the deposition, the mask is rotating to gradually shelter the substrate.

The resulting gradient thickness FeSe was then patterned into 10 pieces to perform *ex-situ* electrical transport measurements as seen in the inset of Fig. 6(a). The spatially dependent film thickness, varying from 28 nm to 280 nm as designed, was checked by both x-ray reflection (XRR) and scanning electron microscopy (SEM). All the narrow strips cut from the sample show a sharp superconducting transition as seen in Fig. 6(a). With increasing thickness, the normal state resistance decreases monotonically as expected. The zero-resistance transition temperatures ($T_{c0}$) of the gradient thickness sample at different regions vary slightly between 8 K and 10 K (Fig. 6(b)). The $T_{c0}$ values are comparable to those films grown by conventional pulsed laser deposition (PLD) method in our previous work[46], verifying that the synthesis part of this system is highly controllable and stable.

### B.    STM performance

Before carrying out STM/STS experiments, we tested the low temperature performance of the system. The sample holder in the STM scanner head can be cooled down to $\sim 5.5$ K in 3 hrs after the L-He vessel has been precooled with liquid nitrogen. The holding time of liquid helium is around 44 hrs. The average lateral and vertical drifts of the scanner head are 33 pm/hr and 42 pm/hr, respectively. The frequency spectra of the background current at room temperature are shown in Fig. 7(a), in cases of tip retracted, tip approached with feedback turned on or off. The highest peak is lower than $1 \mathrm{pA}/\sqrt{\mathrm{Hz}}$ in the situation of tip approached and feedback turned off. These current frequency spectra clearly demonstrate the effective electrical grounding and vibration isolation of the characterization part of our system. To test the location precision of the sample stage, we found an easily identifiable region on the Au (111) surface as shown in Fig. 7(b). A feature point is marked in the image. Then, the tip is retracted 0.37 μm, and the sample is moved 2 mm along the y-axis and moved back later. The same feature can be observed as seen in Fig. 7(c). The coordinates of the marked point can be calculated by summing up the feedback values of Attocube nano-positioner and Nanonis controller. According to the marked point coordinates before and after moving, the location precision is about 0.3 μm.

The spatial resolution and image quality of our system can be judged from several experiments. A Pt/Ir tip is used for all the tests. We are able to obtain atomic-resolution images of a HOPG cleaved surface at both low temperature and room temperature, as shown in Fig. 8(a). In addition, the herringbone structure and the atomic construction of Au (111) have been clearly resolved after cleaning the sample *in-situ* for several cycles by ion gun sputtering and annealing (Fig. 8(b)). To further verify the stability, we performed STM/STS measurements on a BSCCO single crystal, which is an ideal platform for the test[35]. The sample is cleaved mechanically in the preparation chamber, and then transferred to the STM chamber within 5 minutes. Typical supermodulation images and *dI/dV* spectra are shown in Fig. 8(c) and (d), respectively. During the differential conductance measurement, the tip position is fixed with the tunneling parameters $I_t$ = 50 pA, $V_b$ = -200 mV. The amplitude and frequency of the modulation bias are 1 mV and 734 Hz, respectively. From these results, we can estimate a supermodulation period of around 3 nm and a superconducting gap of 44 meV at 77 K, in agreement well with previous reports[47, 48]. As a final test, a FeSe film deposited in Combi-LMBE chamber is transferred into STM chamber for characterization. The surface topography and atomic resolution images of the FeSe film are shown in Fig. 8(e) and (f). The in-plane lattice constant measured from the atomic resolution image is close to the results from XRD measurements (a = 3.74 Å)[46].

## IV. CONCLUSION

We report the design, assembly and performance of an advanced Combi-LMBE-STM facility, intended to accelerate the materials research by integrating high-throughput film synthesis, *in-situ* surface morphology and electronic states characterization. Compared to traditional film deposition techniques, combi-films can be deposited with parameters like chemical substitution and sample thickness varied continuously in a single run. A rotary mask is used in our system to reduce the accumulative error from conventional reciprocating motion. After the deposition, the samples are transferred to the specially designed STM chamber for *in-situ* characterization of surface morphology and electronic states. For compatibility with the combi-film, a large range closed-loop XY coarse positioning module was developed to realize motion in a range of 15 mm × 15 mm but with sub-micro meter precision. With this feature, we can measure the evolution of surface morphology and electronic states as a function of varying parameter (composition or thickness) with high efficiency. The performance of the whole system is demonstrated by the growth of high-quality FeSe film with thickness gradient, and topographic images of various samples including HOPG, Au (111), BSCCO single crystal and FeSe thin film. Clean current noise spectra of the tunneling junction and *dI/dV* spectrum of BSCCO have also been obtained. We expect that this system will be ideal for a systematic research on superconducting materials.

## ACKNOWLEDGMENTS

The authors thank Z. B. Wu, Z. Y. Gao, W. H. Wang, X. Y. Hou, R. S. Ma, Y. Li, X. S. Zhu, Prof. L.

Shan, X. Huang, A. W. Wang, J. H. Yan, X. Y. Chen, Y. Q. Xing, H. Yang, X. C. Huang, H. S. Yu, X. Zhang, W. Hu, Y. L. Jia, Y. J. Shi, X. J. Wei, M. Y. Qin, Z. F. Lin, D. Li, X. Y. Jiang, J. S. Zhang, Z. Y. Zhao and Q. Li for useful discussions and helps. We also thank Prof. W. A. Hofer, Prof. S. X. Du and Prof. A. V. Silhanek for polishing the manuscript. This work was supported by the Scientific Instruments and Equipment Project of Chinese Academy of Sciences (YZ201450), the Special Fund for Research on National Major Research Instruments of NSFC (11927808), CAS Key Technology Research and Development Team Project (GJJSTD20170006), the Huairou Science Center of Beijing Municipal Science and Technology Project (Z181100003818013), the National Key Basic Research Program of China (2015CB921000, 2016YFA0300301, 2017YFA0302902, 2017YFA0303003 and 2018YFB0704102), the National Natural Science Foundation of China (11674374, 11834016 and 11804378), the Strategic Priority Research Program of Chinese Academy of Sciences (XDB25000000), the Key Research Program of Frontier Sciences, CAS (QYZDB-SSW-SLH008 and QYZDY-SSW-SLH001), CAS Interdisciplinary Innovation Team, Beijing Natural Science Foundation (Z190008)

### References

[1] P. A. Grunberg Rev. Mod. Phys. **80,** 1531 (2008).

[2] M. Imada, A. Fujimori and Y. Tokura Rev. Mod. Phys. **70,** 1039 (1998).

[3] K. Jin, N. P. Butch, K. Kirshenbaum, J. Paglione and R. L. Greene Nature **476,** 73 (2011).

[4] B. Keimer, S. A. Kivelson, M. R. Norman, S. Uchida and J. Zaanen Nature **518,** 179 (2015).

[5] L. D. Pham, T. Park, S. Maquilon, J. D. Thompson and Z. Fisk Phys. Rev. Lett. **97,** (2006).

[6] F. Kretzschmar, T. Bohm, U. Karahasanovic, B. Muschler, A. Baum, D. Jost, J. Schmalian, S. Caprara, M. Grilli, C. Di Castro, J. G. Analytis, J. H. Chu, I. R. Fisher and R. Hackl Nat. Phys. **12,** 560 (2016).

[7] M. Vojta Rep. Prog. Phys. **66,** 2069 (2003).

[8] S. Sachdev Physica C **470,** S4 (2010).

[9] N. P. Butch, K. Jin, K. Kirshenbaum, R. L. Greene and J. Paglione P. Natl. Acad. Sci. USA **109,** 8440 (2012).

[10] J. J. Hanak J. Mater. Sci. **5,** 964 (1970).

[11] K. Jin, R. Suchoski, S. Fackler, Y. Zhang, X. Q. Pan, R. L. Greene and I. Takeuchi APL Mater. **1,** 042101 (2013).

[12] J. Wu, O. Pelleg, G. Logvenov, A. T. Bollinger, Y. J. Sun, G. S. Boebinger, M. Vanevic, Z. Radovic and I. Bozovic Nat. Mater. **12,** 877 (2013).

[13] H. Koinuma and I. Takeuchi Nat. Mater. **3,** 429 (2004).

[14] K. Kennedy, Stefansk.T, G. Davy, V. F. Zackay and E. R. Parker J. Appl. Phys. **36,** 3808 (1965).

[15] X. D. Xiang, X. D. Sun, G. Briceno, Y. L. Lou, K. A. Wang, H. Y. Chang, W. G. Wallacefreedman, S. W. Chen and P. G. Schultz Science **268,** 1738 (1995).

[16] Z. L. Luo, B. Geng, J. Bao and C. Gao J. Comb. Chem. **7,** 942 (2005).

[17] P. Ahmet, Y. Z. Yoo, K. Hasegawa, H. Koinuma and T. Chikyow Appl. Phys. A-Mater. Sci. Process **79,** 837 (2004).

[18] M. Murakami, K. S. Chang, M. A. Aronova, C. L. Lin, M. H. Yu, J. H. Simpers, M. Wuttig, I. Takeuchi, C. Gao, B. Hu, S. E. Lofland, L. A. Knauss and L. A. Bendersky Appl. Phys. Lett. **87,** 112901 (2005).

[19] T. Byrne, L. Lohstreter, M. J. Filiaggi, Z. Bai and J. R. Dahn Surf. Sci. **602,** 2927 (2008).

[20] J. E. Daniels, W. Jo, J. Rodel, V. Honkimaki and J. L. Jones Acta. Mater. **58,** 2103 (2010).

[21] J. L. Jones, A. Pramanick and J. E. Daniels Appl. Phys. Lett. **93,** 152904 (2008).

[22] Z. L. Luo, B. Geng, J. Bao, C. H. Liu, W. H. Liu, C. Gao, Z. G. Liu and X. L. Ding Rev. Sci. Instrum. **76,** 095105 (2005).

[23] J. S. Wang, Y. Yoo, C. Gao, I. Takeuchi, X. D. Sun, H. Y. Chang, X. D. Xiang and P. G. Schultz Science **279,** 1712 (1998).

[24] Z. W. Jin, T. Fukumura, M. Kawasaki, K. Ando, H. Saito, T. Sekiguchi, Y. Z. Yoo, M. Murakami, Y. Matsumoto, T. Hasegawa and H. Koinuma Appl. Phys. Lett. **78,** 3824 (2001).

[25] Y. Matsumoto, M. Murakami, T. Shono, T. Hasegawa, T. Fukumura, M. Kawasaki, P. Ahmet, T. Chikyow, S. Koshihara and H. Koinuma Science **291,** 854 (2001).

[26] P. J. Curran, H. A. Mohammed, S. J. Bending, A. E. Koshelev, Y. Tsuchiya and T. Tamegai Sci. Rep. **8,** 2103 (2018).

[27] A. M. Chang, H. D. Hallen, L. Harriott, H. F. Hess, H. L. Kao, J. Kwo, R. E. Miller, R. Wolfe, J. Vanderziel and T. Y. Chang Appl. Phys. Lett. **61,** 1974 (1992).

[28] T. Wei, X. D. Xiang, W. G. WallaceFreedman and P. G. Schultz Appl. Phys. Lett. **68,** 3506 (1996).

[29] K. Hasegawa, P. Ahmet, N. Okazaki, T. Hasegawa, K. Fujimoto, M. Watanabe, T. Chikyow and H. Koinuma Appl. Surf. Sci. **223,** 229 (2004).

[30] K. Lai, M. B. Ji, N. Leindecker, M. A. Kelly and Z. X. Shen Rev. Sci. Instrum. **78,** 063702 (2007).

[31] K. C. Hewitt, P. A. Casey, R. J. Sanderson, M. A. White and R. Sun Rev. Sci. Instrum. **76,** 093906 (2005).

[32] B. C. Stipe, M. A. Rezaei and W. Ho Science **280,** 1732 (1998).

[33] A. J. Heinrich, J. A. Gupta, C. P. Lutz and D. M. Eigler Science **306,** 466 (2004).

[34] V. Madhavan, W. Chen, T. Jamneala, M. F. Crommie and N. S. Wingreen Science **280,** 567 (1998).

[35] Ø. Fischer, M. Kugler, I. Maggio-Aprile, C. Berthod and C. Renner Rev. Mod. Phys. **79,** 353 (2007).

[36] J. D. Hackley, D. A. Kislitsyn, D. K. Beaman, S. Ulrich and G. V. Nazin Rev. Sci. Instrum. **85,** 103704 (2014).

[37] B. C. Stipe, M. A. Rezaei and W. Ho Rev. Sci. Instrum. **70,** 137 (1999).

[38] J. W. Lyding, S. Skala, J. S. Hubacek, R. Brockenbrough and G. Gammie Rev. Sci. Instrum. **59,** 1897 (1988).

[39] G. Meyer Rev. Sci. Instrum. **67,** 2960 (1996).

[40] B. G. Orr, C. W. Snyder and M. Johnson Rev. Sci. Instrum. **62,** 1400 (1991).

[41] W. Z. Lin, A. Foley, K. Alam, K. K. Wang, Y. H. Liu, T. J. Chen, J. Pak and A. R. Smith Rev. Sci. Instrum. **85,** (2014).

[42] Z. B. Wu, Z. Y. Gao, X. Y. Chen, Y. Q. Xing, H. Yang, G. Li, R. S. Ma, A. W. Wang, J. H. Yan, C. M. Shen, S. X. Du, Q. Huan and H. J. Gao Rev. Sci. Instrum. **89,** 113705 (2018).

[43] P. E. Dyer, A. Issa, P. H. Key and P. Monk Supercond. Sci. Tech. **3,** 472 (1990).

[44] R. C. Estler, N. S. Nogar, R. E. Muenchausen, X. D. Wu, S. Foltyn and A. R. Garcia Rev. Sci. Instrum. **62,** 437 (1991).

[45] H. S. Yu, J. Yuan, B. Y. Zhu and K. Jin Sci. China-Phys. Mech. Astron. **60,** 087421 (2017).

[46] Z. P. Feng, J. Yuan, G. He, W. Hu, Z. F. Lin, D. Li, X. Y. Jiang, Y. L. Huang, S. L. Ni, J. Li, B. Y. Zhu, X. L. Dong, F. Zhou, H. B. Wang, Z. X. Zhao and K. Jin Sci. Rep. **8,** 4039 (2018).

[47] S. H. Pan, J. P. O'Neal, R. L. Badzey, C. Chamon, H. Ding, J. R. Engelbrecht, Z. Wang, H. Eisaki, S. Uchida, A. K. Guptak, K. W. Ng, E. W. Hudson, K. M. Lang and J. C. Davis Nature **413,** 282 (2001).

[48] C. Renner, B. Revaz, J. Y. Genoud, K. Kadowaki and O. Fischer Phys. Rev. Lett. **80,** 149 (1998).

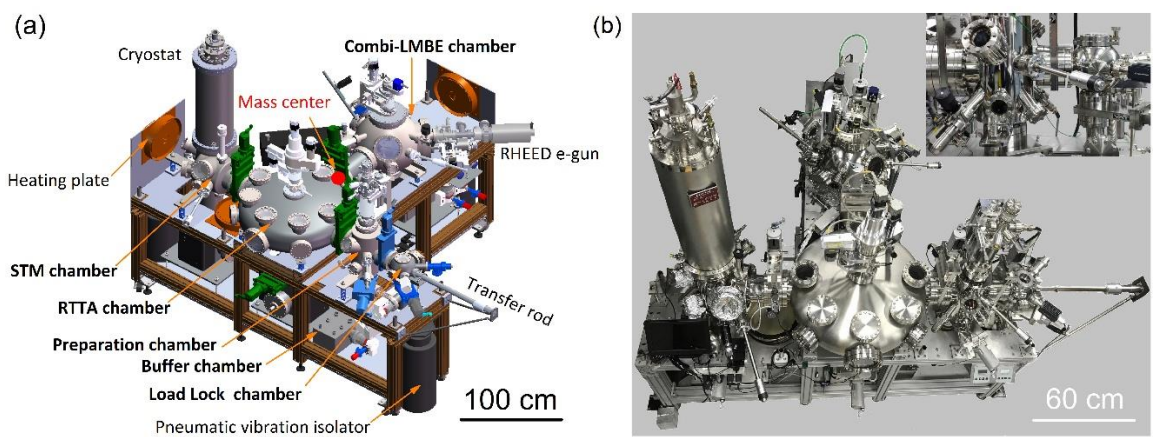

FIG. 1. **3D CAD drawing and photograph of the system.** (a) Isometric view of the system. The chambers, transfer rod, vibration isolators, cryostat, RHEED, etc. are indicated in the figure. The red dot shows the mass center of the system. (b) Photograph of the system. The preparation chamber and the load-lock chamber are shown in the inset.

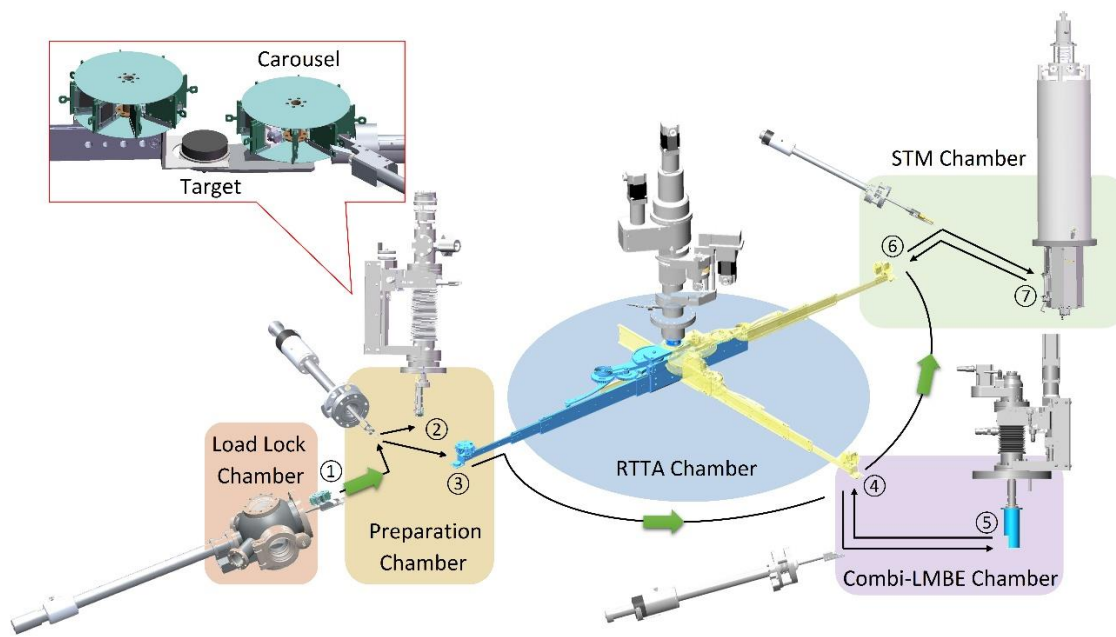

FIG.2 **Design of sample transfer.** The sample/substrate/target/tip are loaded from load-lock chamber. Samples and tips can be transferred to preparation chamber by the transfer rod ①. A wobble stick mounted on the preparation chamber grabs the sample holder and place it to ② for ion sputtering and annealing. The sample/substrate/target/tip are transferred to RTTA at position ③. The target is transferred by a fork as shown in inset. Then, the target is transferred from the RTTA chamber to the Combi-LMBE chamber ④ for deposition. The substrate is positioned at the sample stage ⑤. The samples and tips are transferred from RTTA ⑥ to the STM chamber ⑦ for characterization.

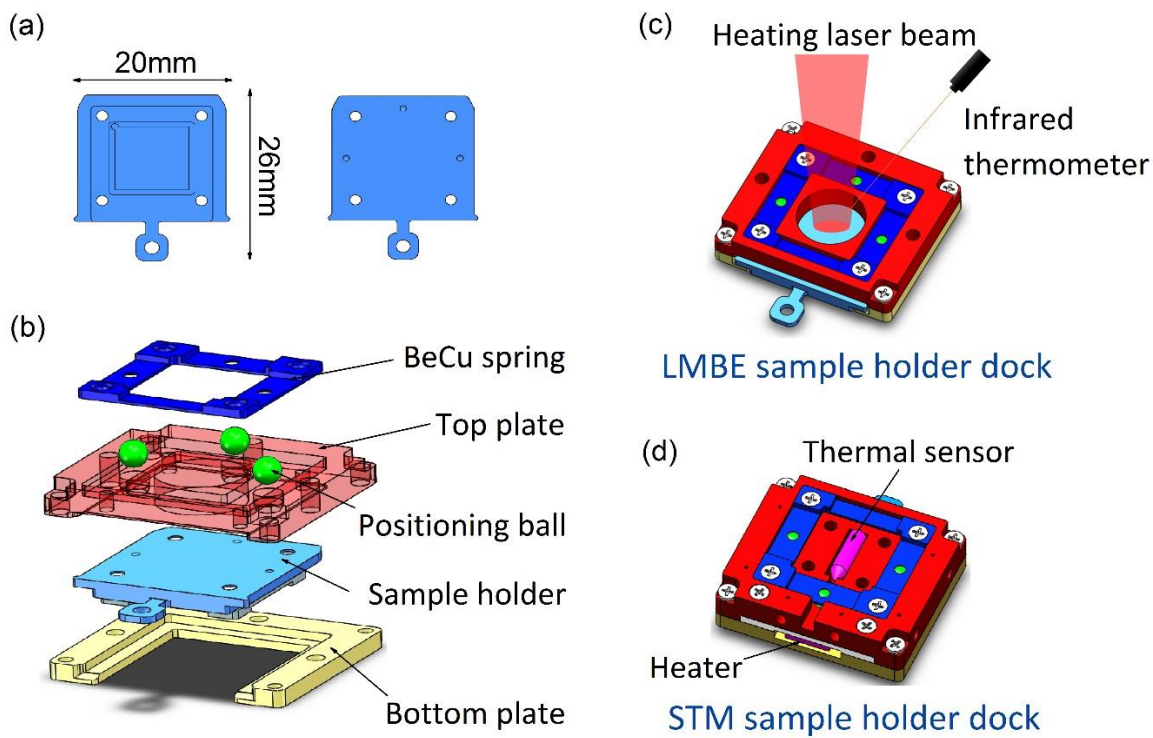

FIG. 3. **Design of the sample holder.** (a) Top and bottom view of the sample holder. (b) Explosion view of the sample holder dock. (c) Sample holder dock for Combi-LMBE chamber. The red transparent area shows the heating laser optical path. An infrared thermometer focuses on the back of the sample holder at an angle of 45 degrees. (d) Sample holder dock for STM chamber. The thermal sensor and heater are located in the back and side of sample holder dock respectively.

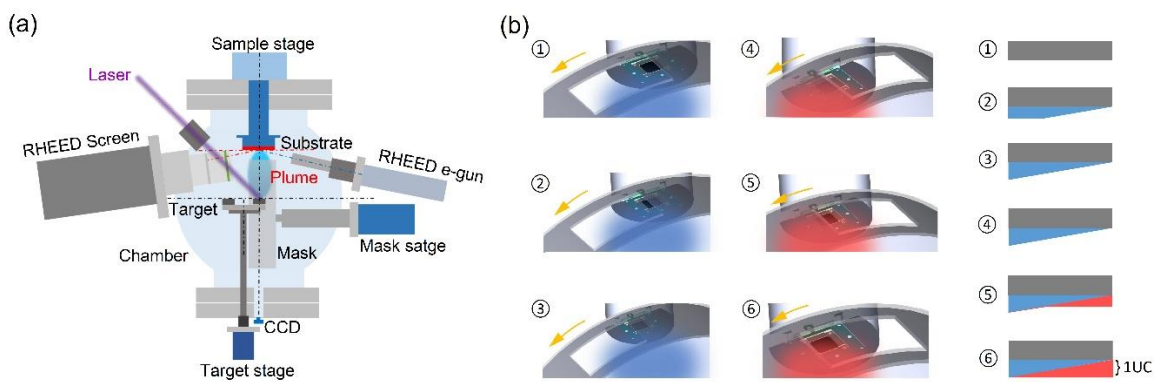

FIG. 4. **Design of the Combi-LMBE unit and schematic of the combi-film deposition.** (a) Distribution of the main components in the Combi-LMBE chamber. (b) Schematic diagram of the combi-film deposition stages. The left and middle columns show the process of deposition in one cycle. The right column shows the corresponding outcome during sample deposition, with the substrate in grey, the component A in blue, and the component B in red.

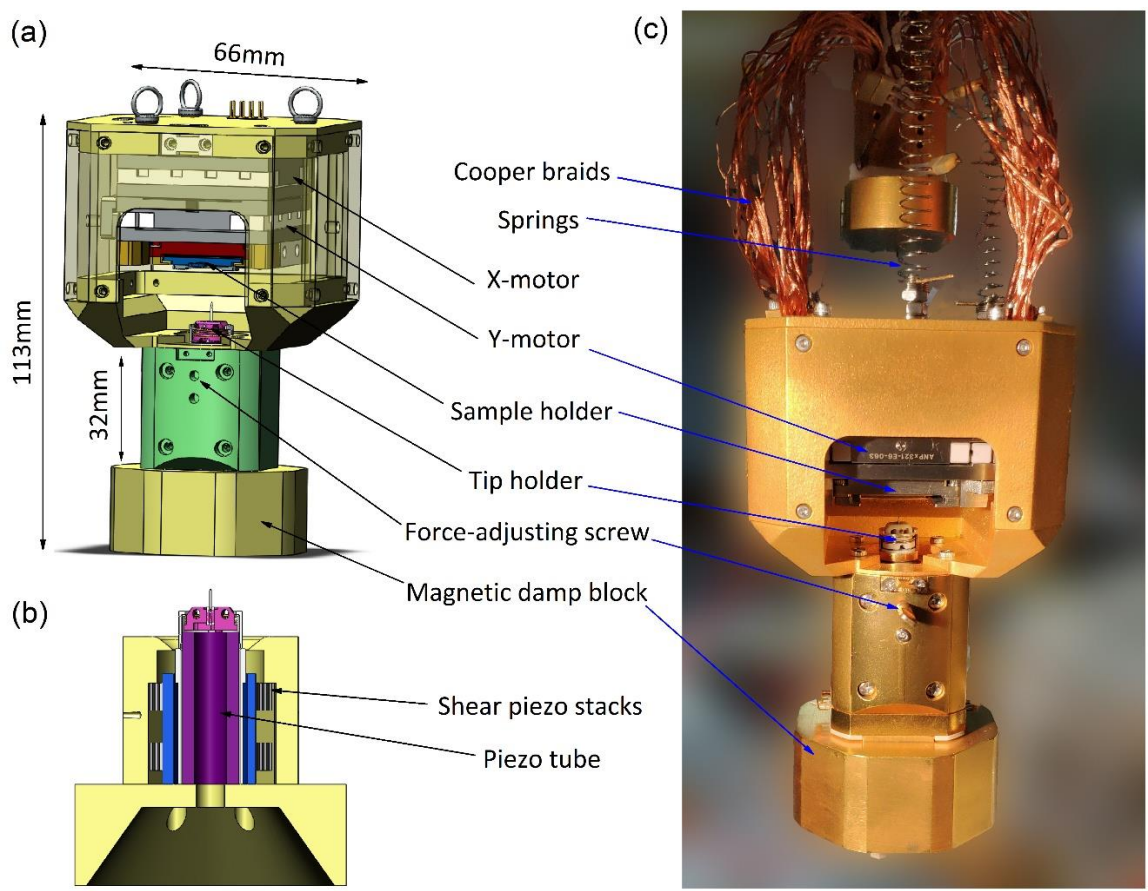

FIG. 5. **STM scanner head.** (a) The 3D model of a home-made scanner head. The sample holder dock is mounted in the X- and Y-motor working in a range of 15 mm ×15 mm. (b) A cross-section of the Z-motor. (c) Photograph of the STM scanner head.

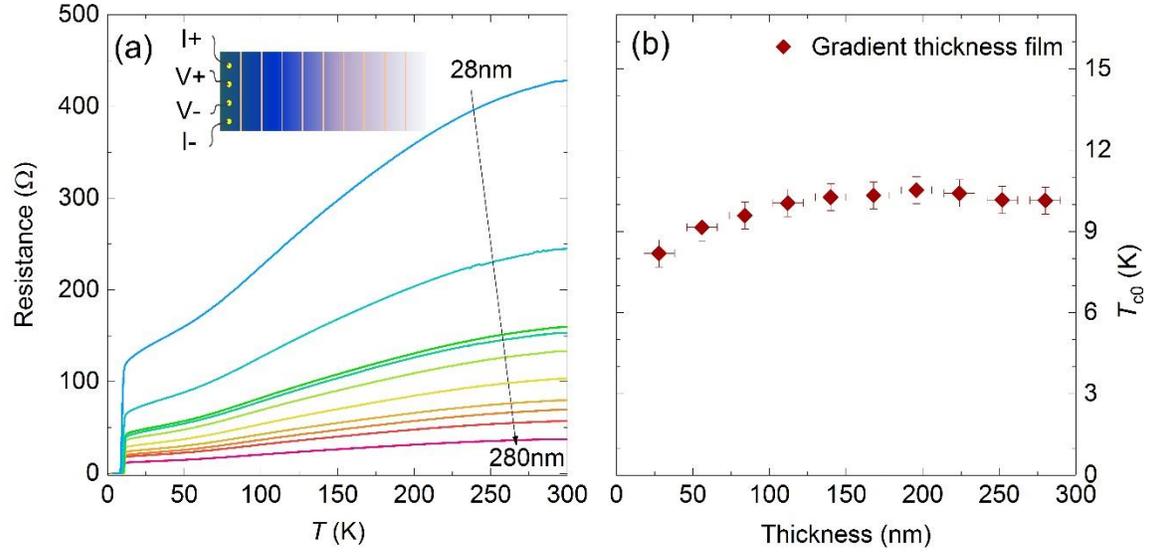

FIG. 6. **The transport properties of FeSe film with thickness gradient.** (a) Temperature dependence of resistance of FeSe film with gradient thickness at different regions. The thickness is the average value at the local region. Inset: Electrical transport measurement configuration of the gradient thickness film, and orange lines show the dicing of the sample into ten pieces. (b) Thickness dependence of $T_{c0}$ for a gradient thickness film.

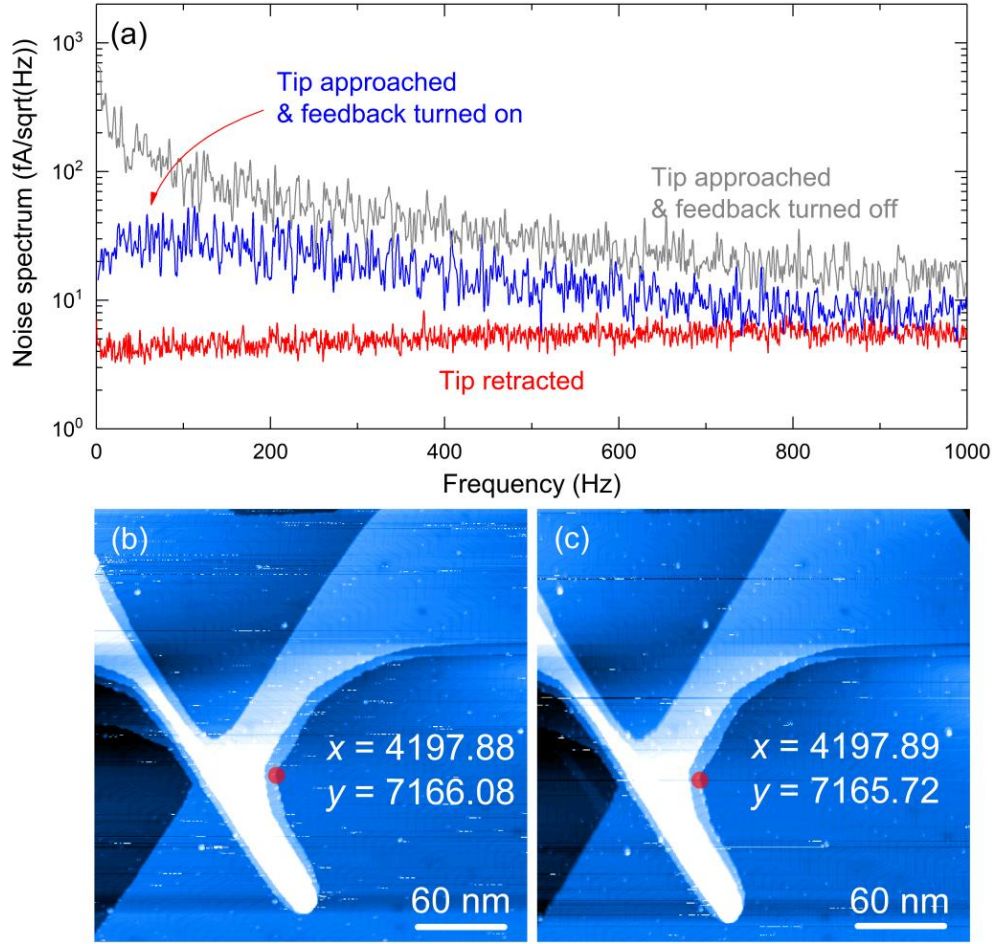

FIG. 7. **The noise spectra of tunneling junction and location precision test.** (a) The noise spectra of the background current with the tip retracted (red line), tip approached with feedback turned on (blue line), and tip approached with feedback turned off (grey line). Before the acquisition of the spectra with tip approached, the tip is positioned at a fixed height with $I_t$ = 0.3 nA, $V_b$ = -0.5V. All the spectra are acquired on Au(111) surface at room temperature. [(b) and (c)] The STM images on Au(111) surface at the same region before (b) and after (c) moving the sample away and moving it back. The coordinates with the unit of micro-meter are shown in the images. The coordinates of the marked point are calculated by summing up the feedback values of Attocube nano-positioner and Nanonis controller.

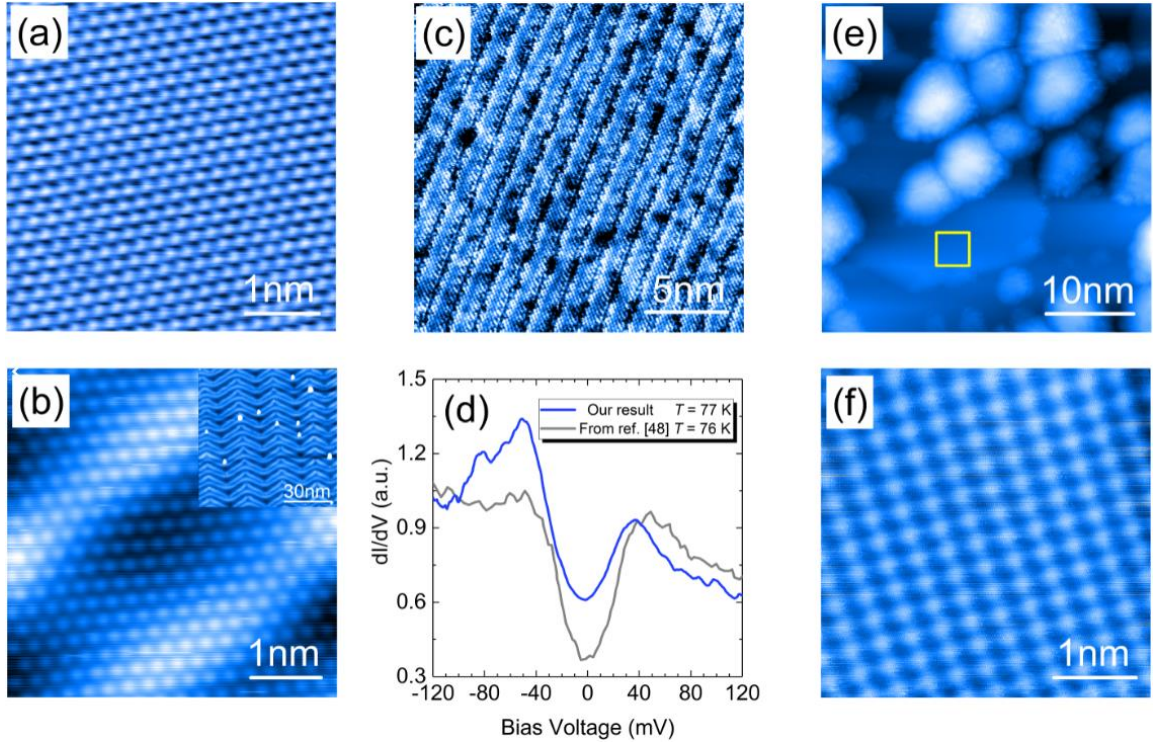

FIG. 8. **Characterization of STM performance.** (a) Atomic construction of the HOPG surface, $I_t$ = 0.1 nA, $V_b$ = -0.5 V, $T$ = 77 K. (b) Atomic construction of the Au (111) surface, $I_t$ = 0.25 nA, $V_b$ = -0.5 V, $T$ = 5.5 K. Inset: Herringbone reconstruction of the Au (111) surface, $I_t$ = 0.1 nA, $V_b$ = -0.5 V, $T$ = 77 K. (c) Supermodulation of an *in-situ* cleaved clean surface in BSCCO single crystal. $I_t$ = 0.05 nA, $V_b$ = -0.2 V, $T$ = 5.5 K. (d) $dI/dV$ spectrum of BSCCO single crystal with $I_t$ = 0.05 nA and $V_b$ = -0.2 V (blue line). The amplitude and frequency of the modulation bias are 1 mV and 734 Hz, respectively. The gray $dI/dV$ spectrum of BSCCO is adapted from Ref. 48. (e) Topographic images of the FeSe thin film surface, $I_t$ = 0.1 nA, $V_b$ = -2 V, $T$ = 77 K. (f) Atomic construction of the FeSe thin film surface taken from the yellow square region in (e), $I_t$ = 0.6 nA, $V_b$ = -0.5 V, $T$ = 77 K.